\documentclass[12pt]{iopart}

\begin{document}

\title{Breaking symmetry, breaking ground}

\author{Mark Hindmarsh}

\address{Department of Physics \& Astronomy, University of Sussex, Brighton BN1 9QH, UK\\
Department of Physics and Helsinki Institute of Physics, PL 64, 00014 University of Helsinki, Finland}
\ead{m.b.hindmarsh@sussex.ac.uk}
\vspace{10pt}

\vspace{10pt}

\textit{Topology of cosmic domains and strings} \cite{Kibble:1976sj}, written in 1976, 
launched the field of cosmic topological defects, and
was seminal for the enormous expansion in research linking particle physics to cosmology in the last few decades. 
 Topological defects are extended structures in field theories, which exist in continuous ordered media, and are also present in models of particle physics which extend the Standard Model.  In our three dimensions of space, these structures may have their energy concentrated around points, lines, or planes. For example,  superfluid Helium can support linear structures called vortices, around which the fluid circulates without viscosity. A prototype for the Standard Model of particle physics known as the Georgi-Glashow model exhibits spherically symmetric topological defects from which magnetic field lines emerge -- these defects are therefore magnetic monopoles.

The paper showed how to predict what topological defects would form at a symmetry-breaking phase transition 
in the early universe, and also made an estimate of the average separation of the defects.  All one needs to know are the symmetries of the ground state before and after the phase transition: the mathematical theory of topology can then be applied to calculate what structures are allowed.  More specifically, $n$-dimensional objects will form in $d$ space dimensions if a certain topological quantity, the homotopy group of order $d-n-1$, is non-trivial - that is, is larger than the trivial group consisting of the identity element only.

The paper led to some immensely important developments.  It was one of the seeds of the theory of cosmic inflation, as the lack of any evidence for magnetic monopoles in the universe meant that there was a contradiction between the idea of Grand Unification of forces (taken very seriously in the late 1970s) and the standard hot Big Bang.  Inflation -- a period of accelerated expansion -- was partly motivated by its feature of diluting any population of monopoles to unobservably low densities \cite{Guth:1980zm}.  

Kibble's proposal that linear topological defects  -- cosmic strings -- could produce density perturbations was taken up by Zel'dovich \cite{Zeldovich:1980gh} and Vilenkin \cite{Vilenkin:1981iu}, who used strings to construct a theory of galaxy formation. For a while, strings rivalled the currently accepted one based on the quantum fluctuations during inflation. It is interesting that Kibble himself was pessimistic about cosmic strings having any observable consequence, writing ``We can expect only one string within the visible universe, so that looking for cosmic strings directly would be pointless.Ó

However, as Zel'dovich, Vilenkin and many others pointed out, there are ways to look for strings and other topological defects, and the suggestions have been widely taken up, notably by the Planck collaboration in the Cosmic Microwave Background \cite{Ade:2013xla}. Kibble's paper also inspired several experiments in condensed matter physics, for example investigations of the prediction of the spontaneous formation of vortices in Helium during a rapid quench to the superfluid phase \cite{Ruutu:1995qz,hendry1994generation}.  

It turned out that Kibble's estimate of the density of topological defects was incorrect, based as it was on an equilibrium argument.  The true story was uncovered subsequently by Zurek \cite{Zurek:1996sj}, who showed how the density was related to rate at which the system was taken through the phase transition, and made predictions for the scaling law relating them. The investigation of these scaling laws continues to be an active and experimentally challenging research area \cite{delCampo:2013nla}.

One of the most significant theoretical developments since 1976 has been string theory, which proposes that the fundamental objects of nature are strings rather than particles.  It was originally thought that fundamental strings were too massive to extend across the visible universe without exerting strong and easily observable gravitational perturbations \cite{Witten:1985fp}. A deeper understanding of the theory brought with it the realisation that strings' effective mass could be reduced by placing them at a distance in an extra dimension. It was also discovered that there were topological defect-like extended objects known as D-branes, and that D-strings and fundamental strings could join and split to form a complex branching network. Such a network could have been formed in the early universe at the end of a period of inflation \cite{Jones:2002cv,Copeland:2003bj}.

Searches for cosmic strings continue, both through their gravitational effects and their decay into high-energy particles \cite{Copeland:2011dx,Hindmarsh:2011qj}.  One outstanding issue is the proportion of energy flowing into gravitational radiation relative to the particle flux, which has proved difficult to resolve due to the non-linearity of the field equations and the huge scale separation between the width of the string (less than about $10^{-18}$ m) and the average curvature radius (up to about $10^{18}$ m today). It is possible that fundamental cosmic strings decay predominantly into gravitational radiation while strings from phase transitions produce mainly high energy particles, offering a way to distinguish between them. The recent detection of gravitational waves by the LIGO and VIRGO collaborations \cite{Abbott:2016blz} 
promises an exciting new observational window through which cosmic strings may finally be seen.

\vspace{5mm}

\textit{This article is dedicated to Tom, 
who died while it was being prepared. His quiet brilliance and humanity were a great inspiration to me.}

\newpage

\bibliographystyle{JHEP}

\bibliography{CosmicStrings.bib,Inflation.bib}

\end{document}